\documentclass[traditabstract]{aa}
\usepackage{natbib}
\usepackage{subfig}
\usepackage{amsmath}
\usepackage{enumerate}
\def\vec#1{{\bf#1}}

\usepackage{aas_macros}
\usepackage{graphicx}
\usepackage{txfonts}

\begin{document}

\title{Magnetic power spectra from Faraday rotation maps}

\subtitle{\texttt{REALMAF} and its use on Hydra A}

\author{Petr Kuchar \and Torsten En{\ss}lin}

\institute{Max Planck Institute for Astrophysics, Karl-Schwartzschildstr.1, 85741 Garching, Germany}

\abstract{We develop a novel maximum-a-posteriori method to measure magnetic
power spectra from Faraday rotation data and implement it in the
\texttt{REALMAF} code. 
A sophisticated model for the magnetic autocorrelation in real space permits us to alleviate previously required simplifying assumptions in the processing. 
We also introduce a way to treat the divergence relation of the magnetic field with a multiplicative factor in Fourier space, with which we can model the magnetic autocorrelation as a spherically symmetric function. 
Applied to the dataset of Hydra A north, we find a power law power
spectrum on spatial scales between 0.3 kpc and 8 kpc, with no visible
turnover at large scales within this range and a spectral index consistent with a
Kolmogorov-like power law regime. 
The magnetic field strength profile seems to follow the electron density profile with an index
$\alpha=1$. 
A variation of $\alpha$ from 0.5 to 1.5 would lead to a spectral index between 1.55 and 2.05. 
The extrapolated magnetic field strength in the cluster centre highly depends on the assumed projection angle of the jet. 
For an angle of $45^o$ we derive extrapolated $36\,\mu$G in the centre and directly probed $16\,\mu$G at 50 kpc
radius.}


\maketitle

\section{Introduction}
\label{introduction}
\subsection{Motivation}

Intergalactic magnetic fields are of great scientific interest. They are part of the intergalactic metabolism, they affect the propagation of heat and cosmic rays, and they are tracers of the dynamical state of the intergalactic medium.
One way to retrieve their properties is to measure the Faraday rotation effect they cause on traversing polarised light. 
This has lead to significant insight into the presence, strength, and morphology of magnetic fields in galaxy clusters.
%\footnote{}

Although it is not possible to retrieve  the exact magnetic field in position space from Faraday rotation data, the magnetic power spectrum is accessible and contains information on the typical field strength, correlation length, and the distribution of magnetic energy onto different length-scales. 
All these quantities can be predicted in magneto-genesis theories. 
The latter, the spectral energy distribution, is a sort of fingerprint of these theories. 
Magneto-genesis-scenarios in which the energy is injected on large scales by gas stirring typically exhibit Kolmogorov-like power law spectra \citep{2006MNRAS.366.1437S, 2006A&A...453..447E}. 
Whereas scenarios in which the magnetic fields are built up from small to large scales typically exhibit different spectral slopes \citep{2006PhPl...13e6501S}. 
Magnetic field measurements by Faraday rotation therefore have the potential to distinguish between these theories.

\subsection{Faraday rotation}

The plane of polarisation of light traveling to us through a magnetised plasma is
changed by the wavelength-dependent Faraday rotation effect. 
If the Faraday rotating plasma is entirely located between the radio source (at physical depth $z_\mathrm{s}$ and at position $\vec{x}_\perp$ in the plane of sky) and the observer (assumed to be at $z= \infty$), the effect is a plain change of the original polarisation angle $\chi_0$ according to
\begin{equation}\label{RM}
\chi(\vec{x}_\perp, \lambda)= \chi_0+ \lambda^2\,\underbrace{a_0\int_{z_\mathrm{s}}^\infty \!\!\!\! dz\,
n_e(\vec{x}_\perp,z) \, B_z(\vec{x}_\perp,z)}_{=\phi(\vec{x}_\perp)},
\end{equation}
with a proportionality constant $a_0=e^3/(2\pi\, m_e^2c^4)$, with $c$, $e$, and $m_e$ being the speed of light, the electron charge, and the mass, respectively. 
Furthermore, $n_e(\vec{x})$ and $\vec{B}(\vec{x})$ are the electron density and the magnetic field within the Faraday rotating foreground medium, and $\phi(\vec{x}_\perp)$ denotes the Faraday depth of the source at $\vec{x}_\perp$.
Although the original polarisation angle $\chi_0$ is unknown, measurements of the polarisation angle at a range of wavelengths $\lambda$ provide this observationally, and also an estimator of the Faraday depth $\phi$ via a fit to Eq. \ref{RM}. 
The observational estimate of the Faraday depth is called the rotation measure (RM), because it expresses the rate at which the linear polarisation rotates as a function of $\lambda^2$. 

Some of the many studies of magnetic fields via Faraday rotation are \cite{1979AJ.....84..725D},  \cite{1982ApJ...252...81L}, \cite{1987ApJ...316..611D}, \cite{1988Natur.331..149L}, \cite{1989ApJ...347..144H}, \cite{1990ApJ...360...41T}, \cite{1993ApJ...411..518G}, \cite{1993AJ....105..778G}, \cite{1994AJ....108.1523G}, \cite{1995A&A...302..680F}, \cite{1998ApJ...495..564K}, \cite{1998ApJ...495..564K}, \cite{1999A&A...344..472F, 1999A&A...341...29F}, \cite{2001ApJ...547L.111C}, \cite{2001MNRAS.326....2T}, \cite{2002ARA&A..40..319C}, \cite{2002MNRAS.334..769T}, \cite{2002ApJ...567..202E}, \cite{2004A&A...424..429M}, \cite{2006MNRAS.368...48L}, \cite{2006ApJ...637...19X}, \cite{2007MNRAS.382...67T}, \cite{2009ApJ...705L..90C}, \cite{2009ApJ...707..114F}, and \cite{2009MNRAS.393.1073B}, where we omitted works directly addressed in the following.

\subsection{REALMAF}

In this work, we present the code \texttt{REALMAF} (\texttt{REAL MA}gnetic \texttt{F}ields), which is aimed at measuring magnetic power spectra in galaxy clusters from maps of Faraday rotation measurements and apply this code to the Hydra A cluster.

As will be detailed in the following sections, our method relies on these informations and assumptions:
\begin{itemize}
\item the linearity of the measurement process (see Sect. \ref{sec:FS})
\item the maps of rotation measures and their errors (see Sect. \ref{appldata})
\item knowledge of the geometry of the Faraday active volume, which
  also depends on the projection angle $\theta$ of the jet and the cavity size (see Sect. \ref{applwindow})
\item assuming that there is no polarised radio emission inside the Faraday active volume in front of the radio bubble (see Sect. \ref{applwindow})
\item knowledge of the electron density profile $n_e$ of the cluster atmosphere (see Sect. \ref{applwindow})
\item assuming that the magnetic field strength B follows the electron
  density profile like $B \propto n_e^\alpha$ \citep[e.g.][ and see Sect. \ref{correlationmatrix}]{2001A&A...378..777D}
\item assuming Gaussian probability distributions for the magnetic fields and the measurement errors in the rotation measure map (see Sect. \ref{sec:GRF})
\item assuming statistical isotropy of $B$ (see Sect. \ref{sec:M})
\item assuming solenoidal fields  (see Sect. \ref{sec:M})
\item using a Bayesian approach to inference with a scale-independent prior for the power spectrum (see Sect. \ref{sec:statinf})
.
\end{itemize}
We use $H_0=70\,\textrm{km}\,{s}^{-1}\,\textrm{Mpc}^{-1}$,
$\Omega_m=0.3$, $\Omega_\Lambda=0.7$ and a redshift of 0.0538 for
Hydra A \citep{1991trcb.book.....D}, which gives a linear to angular conversion factor $1"=1.047\textrm{kpc}$.
To follow the tradition in cosmic magnetism studies, we use cgs-units in this paper with the energy unit
${1\,\textrm{erg}=10^{-7}\,\textrm{Joule}}$ and magnetic field
strength unit ${1\,\textrm{G}=10^{-4}\,\textrm{T}}$.

\section {Magnetic field statistics in Faraday screens}
\subsection{Measurement equation}\label{sec:FS}

The rotation measure differs from the Faraday depth as defined by Eq. \ref{RM} only by the measurement noise $n$:
\begin{eqnarray}\label{eq:RMdata}
 \mathrm{RM}(\vec{x}_\perp) &=& \phi(\vec{x}_\perp) + n(\vec{x}_\perp)\nonumber \\
&=& \sum_{i\in\{x,y,z\}}\,\int_{z_\mathrm{s}}^\infty \!\!\!\!   \underbrace{dz\,a_0\,
n_e(\vec{x}_\perp,z)\, \delta_{i\,z} }_{R_{\vec{x}_\perp \, (\vec{x},i)}}\, B_i(\vec{x}_\perp,z)
+ n(\vec{x}_\perp).
\end{eqnarray}

We assume the signal-to-noise ratio of the polarised flux to be sufficiently high for an accurate RM determination within an area $A$ in the sky. 
The n-tuple of the RM values there then forms our data vector $d=(\mathrm{RM}(\vec{x}_{\perp}))_{\vec{x}_\perp\in A}$, from which all deductions will be made.
The RM data can be regarded as the result of a functional linear response operator $R$ acting on the three-dimensional magnetic field configuration and producing a two-dimensional sky image of it,
\begin{equation}\label{eq:d=RB+n}
d = R\, B + n,
\end{equation}
where the elements of the response matrix $R$ are defined in Eq. \ref{eq:RMdata} and the application of $R$ on $B$ should be read as the index-sum and line-of-sight integration given in the same equation.
The extended Faraday effect in the intra-cluster medium in front of the Hydra A radio galaxy was accurately measured by \cite{1993ApJ...416..554T}. 
The intra-cluster medium has a roughly spherical symmetric electron
density $n_e(\vec{x}_\perp,z)$ centred on the Hydra A cD galaxy.
The radio jets of Hydra A inflate bubbles with relativistic, radio emitting plasma, which are visible as cavities in the X-ray emission of the surrounding hot cluster atmosphere \citep{2007ApJ...659.1153W}. 
On its way to the observer the polarised radio emission of these bubbles experiences a rotation of its polarisation plane because of the Faraday effect of the magnetised intra-cluster medium.

\subsection{Gaussian random fields}\label{sec:GRF}

Owing to the information lost in the projection, Eqs. \ref{eq:RMdata} and \ref{eq:d=RB+n} are not
directly invertible. It is therefore impossible to calculate the line-of-sight magnetic
field $B_z$ from $\phi$  and the known electron density profile. 
However, we will see that assuming statistical isotropy of the magnetic fields and using a
statistical approach we are able to retrieve the amplitude spectrum of
the magnetic field in Fourier space, the magnetic power spectrum
$\omega (k)= \langle \vec{B}(\vec{k}) \cdot \overline{\vec{B}}(\vec{k})\rangle$, and
the typical field strength $B$ in position space and its
autocorrelation length $\lambda_B$, which are determined by $\omega (k)$. 
The expectation brackets should be understood as an average over the ensemble of possible magnetic field configurations produced by typical intra-cluster dynamics. 
We will assume that the magnetic field statistics generated by those is Gaussian. 

Gaussian magnetic fields statistics is not what numerical simulations of MHD 
turbulence find, see \citet{2009MNRAS.398.1970W}. However, they are the starting point 
of any analysis of correlation functions. 
If all we know statistically about the fields is their two-point correlation, the assumption 
which only expresses this knowledge is a Gaussian of which this correlation 
matrix is the covariance matrix. Any other distribution function would 
contain more information in a Shannon-Boltzmann sense.

Before we can gain additional information on higher order statistics, we 
need to understand more the two point correlation function, and
this is possible even with an approximate assumption on the statistics.
Non-Gaussianities could modify the result of a power spectrum measurement 
based on the assumption of Gaussian fields. However, this modification can be 
expected to be of a modest nature, much smaller than the systematic 
uncertainties we deal with because of the imperfect modeling of the galaxy cluster's atmosphere structure and the geometry of the radio source that was used to probe for Faraday rotation.
Our method is relatively robust to non-Gaussian modifications, because 
the observational quantities it exploits, the product of RM values at 
different locations, are a linear transformation of the magnetic two-point 
correlation, irrespective of whether the underlying magnetic field statistics are Gaussian or not.

Finally, many other methods also use Gaussian random fields or assume them implicitly, e.g. \citet{2004A&A...424..429M,
2010A&A...513A..30B, 2010arXiv1009.1233B, 2010A&A...514A..71V, 2010A&A...522A.105G}.

Using again the functional vector notation, we can write the Gaussian probability to find a magnetic field configuration $B$ given a  magnetic covariance matrix $M = \langle B \, B^\mathrm{T} \rangle $, with elements $M_{ij}(\vec{x},\vec{y}) = \langle B_i(\vec{x}) \, B_j(\vec{y}) \rangle $:
\begin{equation}\label{eq:GaussB}
 P(B|M) = \mathcal{G}(B,M) =  |2\pi\, M|^{-\frac{1}{2}} \, \exp\left( -\frac{1}{2} \, B^\mathrm{T} M^{-1} \, B \right),
\end{equation}
with $|M|$ being the determinant of the magnetic covariance matrix.

\subsection{Magnetic correlation tensor}\label{sec:M}

If there is a statistical translation invariance, $M_{ij}(\vec{x},\vec{y})$ depends only on the  distance $\vec{x} -\vec{y}$
because the absolute position cannot matter and only relative positions determine the two-point statistics.
As a consequence, the magnetic covariance matrix becomes diagonal in its Fourier space representation with respect to the $\vec{k}$-vector:
\begin{equation}
 \langle \hat{B}_i(\vec{k}) \, \overline{\hat{B}_j(\vec{k'})} \rangle = (2\pi)^3\, \delta(k-k')\, \hat{M}_{ij}(\vec{k}).
\end{equation}
If we deal with statistical isotropy, one moreover finds for solenoidal fields ($\vec{\nabla} \cdot \vec{B}=0 $)
\begin{equation}\label{eq:Mij}
\hat{M}_{ij}(\vec{k}) = \frac{ \omega(k)}{2} \,\left(1-\frac{k_i\, k_j}{k^2} \right) + i\, \varepsilon_{ijl}\, \frac{k_l}{k}\, H(k),
\end{equation}
(see e.g. \cite{1999PhRvL..83.2957S}, \cite{2003A&A...401..835E}) and therefore
\begin{equation}
  \omega(k) = \sum_i \, \hat{M}_{ii}(\vec{k}).
\end{equation}
$H(k)$ specifies the magnetic helicity, which does not play any role in this work, because it does not lead to any detectable signature in Faraday rotation maps alone, as we will see shortly.

\subsection{Faraday statistics}

Assuming Gaussian statistics for turbulent magnetic fields has the pleasant property that the correlation function of the Faraday depth $\phi$ is also Gaussian:
\begin{eqnarray}\label{eq:Gaussphi}
 P(\phi) &=& \int \! \mathcal{D}B \, \delta(\phi- R\, B)\, P(B)\nonumber\\
&=&       \int \! \frac{\mathcal{D}B}{|2\pi\,M|^{\frac{1}{2}}}  \, \int \! \mathcal{D}\eta \exp\left( 2\pi \, i\, \eta^\mathrm{T} (\phi- R\, B)  -\frac{1}{2} \, B^\mathrm{T} M^{-1} \, B \right)
\nonumber\\
&=&   \int \! \mathcal{D}\eta \exp\left( 2\pi \, i\, \eta^\mathrm{T} \phi  -\frac{(2\pi)^2}{2} \, \eta^\mathrm{T} R^\mathrm{T} M \,R\,\eta \right)
\nonumber\\
&=& |2\pi\, R^\mathrm{T} M \,R|^{-\frac{1}{2}} \,\exp\left(   -\frac{1}{2} \, \phi^\mathrm{T} (R^\mathrm{T} M \,R)^{-1}\,\phi \right)\nonumber\\
&=& \mathcal{G}(\phi, R^\mathrm{T} M \,R).
\end{eqnarray}
Here, we used the Fourier representation of the Dirac delta function (in functional spaces) as well as the well known property of Gaussian integrals $\int\! \mathcal{D}\psi \, \exp(-\psi^\mathrm{T} A^{-1}\, \psi/2 + j^\mathrm{T} \psi) = |2\pi A|^{1/2} \,\exp(j^\mathrm{T} A\, j/2)$, which also holds for functional or path integrals denoted by $\int\! \mathcal{D}\psi$. 
The $\phi$-covariance $C_\phi = \langle \phi \, \phi^\mathrm{T} \rangle =  R^\mathrm{T} M \,R$ reads for our specific response and magnetic covariance matrices
\begin{equation} \label{eq:Cphi}
  \begin{aligned}
C_\phi(\vec{x}_\perp,\vec{x}'_\perp) = 
a_0^2&\int_{z_1}^\infty \!\!\! dz \int_{z_2}^\infty \!\!\! dz' \, n_e(\vec{x}) \,n_e(\vec{x}') \,
\underbrace{
\langle\,
  B_z(\vec{x}) 
 B_z(\vec{x}')\,\rangle}_{M_{zz}(\vec{x}, \vec{x}')}\, ,
 \end{aligned}
\end{equation}
where $z_1$ and $z_2$ denote the depth to the surface of the radio source at sky position $\vec{x}_\perp$ and $\vec{x}_\perp'$, respectively.  
If the spatial electron distribution is constant, the magnetic fields are statistically homogeneous and isotropic and the observation samples a virtually infinitely extended Faraday active volume, this can be simplified even more, and reads in Fourier space
\begin{eqnarray}
 \langle \hat\phi(\vec{k}) \, \overline{\hat\phi(\vec{k}')} \rangle &=& (2 \pi)^3\, \delta(\vec{k}-\vec{k}') \, \hat{C}_\phi(k)\\
\hat{C}_\phi(k) &=& \frac{1}{2}\, a_0^2 \,L_z\, n_\mathrm{e}^2 \,  \omega(k),
\end{eqnarray}
where $L_z$ is the depth of the Faraday screen.
Thus, clearly the magnetic energy, but not the helicity spectrum can be retrieved from Faraday rotation maps. In the more realistic case the geometry is less simple (finite observed volume, structured electron density, and magnetic field strength profiles, etc.). 
This formula has to be replaced by one which takes these effects into account, as derived in \citet{2003A&A...401..835E}. 
Because then the different Fourier modes couple, it can become simpler to again work in the position space, as we will do below (see Section \ref{correlationmatrix}). 

\subsection{\texttt{REALMAF} and other power spectra estimators}

We model the magnetic power spectrum as a combination of spectral basis functions
\begin{equation}\label{eq:parametrisation}
\omega(k)=\sum_i s_i w_i(k),
\end{equation}
where the spectral amplitudes form our spectral signal $s = (s_i)$. 
The number of bins, and therefore the number of free parameters of our inference problem, 
will be set by a trade-off between the aim to have high spectral resolution and to have small and uncorrelated vertical error bars in the spectrum.
\footnote{The algorithm described in Sect. in \ref{maximazing} may not converge and may fail in cases where the variance of the spectrum is too large, or, equivalently, the inverse problem is too degenerate. 
For the our data this limits the spectral resolution to a maximum of about 7 bins.}

A probability function for 
$s$ given the data $d$, the posterior $P(s|d)$,  will be set up later in this work in Section \ref{basics}.  
Maximising the posterior with respect to  the parameters in $s$ provides an estimator for $\omega(k)$ expressed as amplitudes of relatively independent frequency bands. 
A similar approach has already been applied to the Hydra~A cluster by \cite{2005A&A...434...67V}.
Expressing the spectral basis functions as analogous basis functions for the magnetic autocorrelation allows us to calculate the RM covariance matrix completely in position space, which makes two rough simplifications in the processing used by \cite{2005A&A...434...67V} unnecessary (see Section \ref{comparison} for further discussion). Moreover we include noise effects into the analysis and model the Faraday screen
more accurately using recently improved measurements of the geometry
of the cavities in the Hydra A cluster by
\cite{2007ApJ...659.1153W}.

The Hydra A cluster has been investigated by \cite{2008MNRAS.391..521L} as
well. They compared simulated rotation measure structure functions
with the observed ones. A comparison with their results will be discussed in Section \ref{comparison}.

Another approach for retrieving magnetic power spectra is the one
proposed by \cite{2004A&A...424..429M}. Assuming a power law power spectrum,
they simulated Faraday rotation maps by using the \texttt{FARADAY} code and compared the observed and synthetic profiles of the dispersion of the RM as fit criterion. In order to find a good fit, \citet{2004A&A...424..429M} assumed a power law to restrict the number of degrees of freedom. This method was also applied e.g. by \cite{2006A&A...460..425G} and \cite{2008A&A...483..699G}. 

%%%%
The \texttt{REALMAF} code presented in this work should alleviate several of the restrictions and assumptions of former approaches. 
Although \texttt{REALMAF} also requires the spectral range to be set up by hand, the posterior statistic calculated by it guides the choice. 
The power spectrum within the chosen range is relatively free to adopt any slope required by the data and permitted by the number of assumed spectral bands.

\section{Inference method}
\label{method}

\subsection{Statistical inference and frequentist estimates}\label{sec:statinf}

In this work concepts from statistical inference will be applied to Faraday rotation data to measure magnetic field properties. Because this is traditionally done via frequentist estimates, we here briefly recall the conceptual differences in the methods. 

In a traditional frequentist data analysis the question is asked how likely the data would have emerged under certain model parameters, in the following called the signal. 
The parameters or signal values with high likelihood are then favoured over the ones with low likelihood. 
The likelihood is often estimated by comparing a statistical summary of the real data with that of simulated mock data sets and by characterising their similarity and difference. 
In our case the Faraday rotation dispersion and empirical correlation functions are typically used as statistical summaries of the RM data. 
A comparison of data and model seems to be transparent, because it happens completely in data space. 
One can easily inspect the mock data by eye and compare them to the real one. 
This is one reason why frequentist methods are popular. 
Another one is their lower computational complexity, because they usually only require calculating the measurement process in a forward fashion.

In statistical inference, a different question is addressed: How probable is a signal or a model given the data and other information sources? This is the reverse from the frequentist question and is actually the more relevant question scientifically. 
Answering it requires some knowledge or assumption on how probable the model was before taking the data. This prior probability, in combination with the likelihood probability, permits us -- using the Bayes theorem -- to construct the posterior probability of a model given the data and any earlier knowledge, which then answers the question. 
Because this deduction takes the reverse causal way of the measurement process, the computational complexity is usually higher than that of a comparable frequentist method.

Thus, the statistical inference (or the Bayesian) approach is more abstract in trying to do deductions in the model or signal space. Its answer is a probability function in model space and not a single estimate. The mean and variance of the posterior are often used to characterise the solution. And no mock data and statistical summaries are usually generated, because the inference inserts the observed data into a likelihood function from which a backward deduction of the model is done. 

Frequentist and Bayesian results often agree. However, there are at least three reasons why they can differ, and we believe that the latter two matter here:
\begin{enumerate}
 \item Bayesian methods specify the prior assumptions explicitely, while those are often implicit in frequentist estimates. Because the prior has some impact on the result, different choices of priors will lead to different results. 
\item 
Bayesian methods are better suited to exploit the full information content of the data. 
The summary statistics used in frequentist analysis often looses a significant amount of the information imprinted in the data. 
In our case, we are investigating the correlation between all pairs of pixels, but a typical frequentist statistics might only look at the variance of the individual pixels or the correlation function from a subset of {\it similar} pixels, and therefore exploits much less of the data complexity.
 \item There are situations where no suitable summary statistics exist to construct an empirical likelihood. In our case, calculating an empirical RM correlation function to compare data with mock data only makes sense if the RM field can be assumed to result from a statistically homogeneous process. 
However, we are facing a very inhomogeneous geometry here because of the highly structured cluster atmosphere, with very few comparable lines of sight. How to obtain a meaningful two-point correlation function from the data in this situation is not clear to us.
\end{enumerate}

Because of the last point, the Bayesian inference performed in this work does not provide common quantities of frequentist analysis like mock data, observed RM correlation function, and the like. 
The RM data are confronted with the theoretical covariance matrix that is expected for a given power spectrum of the magnetic field after a model of the inhomogeneous cluster atmosphere is factored out. 
The power spectrum would, in a statistically homogeneous setting, describe an easily observable RM correlation function, but not in our inhomogeneous case.

The Bayesian approach also extracts statistical information from poorly sampled quantities, while providing the correct large uncertainty resulting from the undersampling. 
For example in our case the magnetic modes on scales of or larger than the observed RM map are clearly undersampled.
The largest modes appear as a single number in the data, the mean RM value of the map. 
Because it is a single number, it is hard to invert this into precise spectral information on large-scale magnetic fluctuations. 
Nevertheless, there is some useful information in this number, especially if it is far from zero. 
In this case, the total magnetic spectral power on large spatial scales should be on a level which makes the observed mean RM value sufficiently likely. 
The use of a Jeffreys prior on all spectral amplitudes enforces in our application that the inversion of poorly constrained modes is made in a very conservative fashion. 
With frequentist methods the usage of the information in such poorly sampled modes is more difficult, because no empirical statistics can be built up from a single observation.

\begin{figure*}[htp]
\begin{center}
a)\includegraphics[width=0.42\textwidth,angle=0]{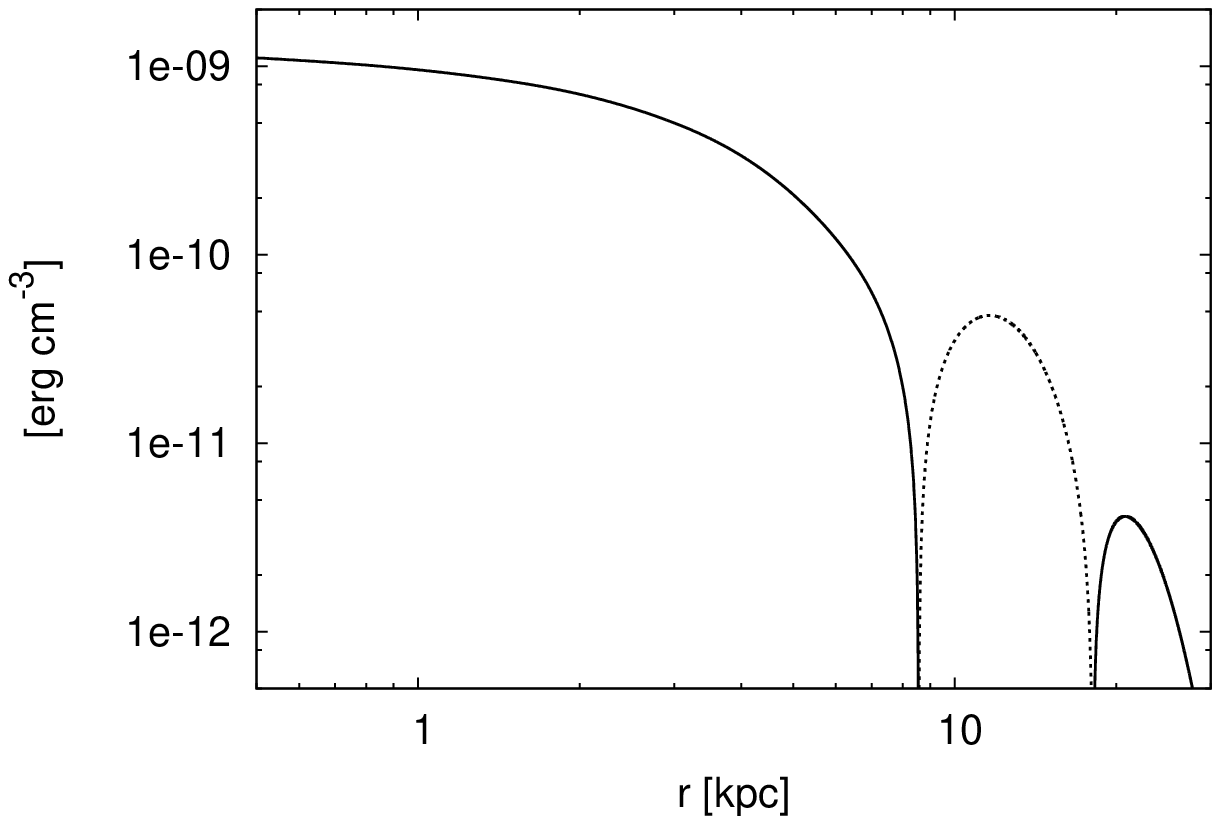}\hspace{50pt}
b)\includegraphics[width=0.42\textwidth,angle=0]{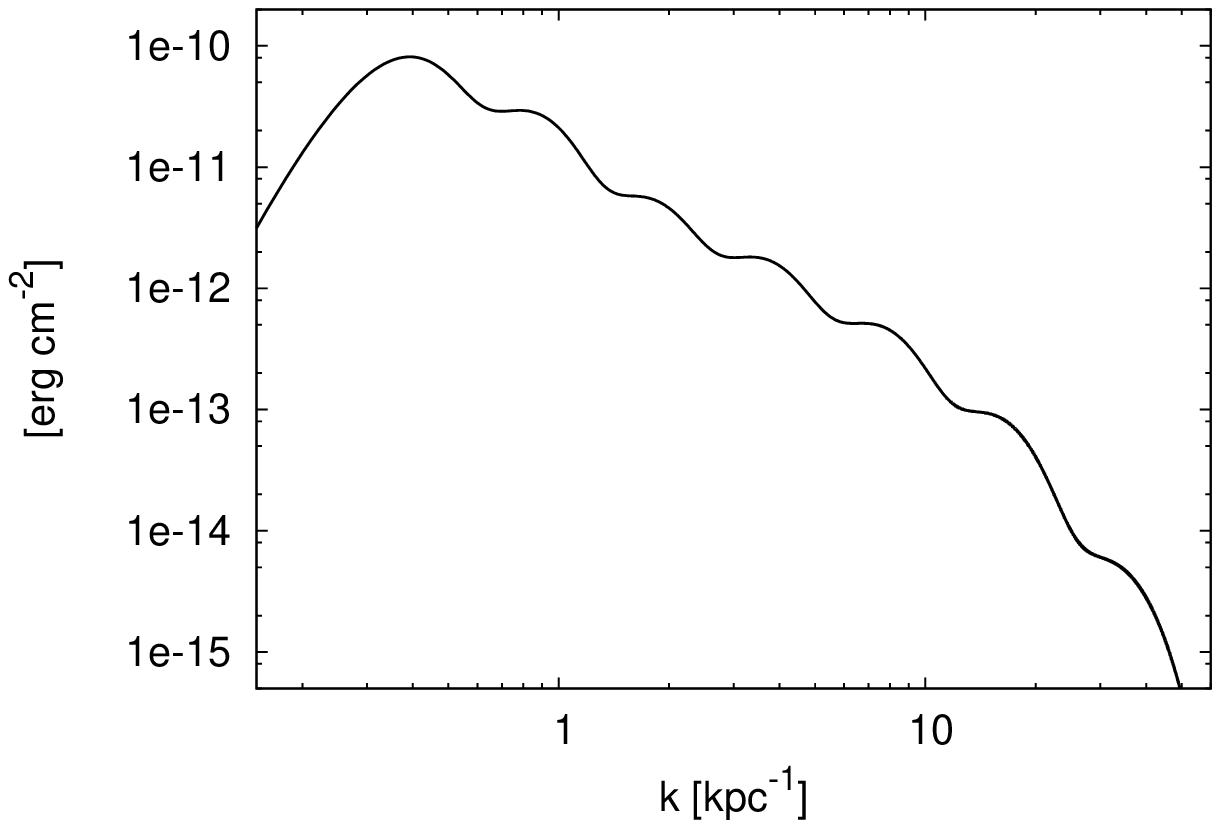}
\caption{Left: the magnetic autocorrelation $\omega(r)$ in position space for a typical magnetic power spectrum reconstruction for the cool core in the Hydra A cluster using seven spectral bins. 
The negative part of $\omega(r)$ is dotted. Right: the counterpart in Fourier space $\omega(k)$. 
The parameterisation can adopt any shape permitted by the seven basis functions seen as wiggles in the power spectrum.
The parameters $\theta=45^o$ and $\alpha=1$ were assumed, with $\theta$ described in Section \ref{applwindow} and $\alpha$ introduced in Eq. \ref{eq:magneticprofile}.}
\label{spectrum}
\end{center}
\end{figure*}

\subsection{Statistical basics}\label{basics}
Our aim is to retrieve from the rotation measures (data $d$) the magnetic power spectrum represented by the spectral amplitudes (signal $s$).
The Bayes theorem expresses the posterior $P(s|d)$ (the conditional probability of the signal given the data) in terms of the likelihood $P(d|s)$ in the following way:
\begin{equation}\label{eq:bayesiantheorem}
P(s|d)=\frac{P(d|s)\, P(s)}{P(d)}.
\end{equation}
Because $P(d)$ is a constant with respect to $s$, the most probable $s$ given $d$ can
be retrieved by maximising the joint probability, which consists of the
likelihood $P(d|s)$ multiplied with the Prior $P(s)$. 
These will be specified now for our problem.

In our case the signal is the ensemble of spectral amplitudes $s = (s_i)_i$. 
We adopt Jeffreys prior for each spectral amplitude, so that $P(s)\propto \prod_i s_i^{-1}$.
Jeffreys prior is a uniform distribution on a logarithmic scale, expressing that we are beforehand not even sure about the magnitude of the magnetic spectrum.

Our likelihood $P(d|s)$ has to incorporate two stochastic processes. 
The first is the realisation of the magnetic field configuration according to Eq. \ref{eq:GaussB}. 
Because $s$ fully determines the here relevant symmetric part of the magnetic correlation tensor, 
$M = M(\omega,H) =M(\sum_i s_i\, \omega_i, H)$, the statistics of the Faraday depth is determined by it as well, 
and it is also Gaussian  according to Eq. \ref{eq:Gaussphi}. 
If the noise is also Gaussian, then the data are Gaussian
with a covariance matrix $C$, with entries ${C_{ij}=\langle d_i\, d_j\rangle}$ for the correlation of RM data at locations $\vec{x}_{\perp , i}$ and  $\vec{x}_{\perp , j}$. 
Using vector and matrix notation for the data vector $d = \phi + n$ and its covariance matrix $C$, we can write the Gaussian as
\begin{equation}\label{eq:likelihood}
P(d|s) = \mathcal{G}(d,C(s)) = 
|2\pi\, C|^{-\frac{1}{2}} \, e^{-\frac{1}{2}d^T C^{-1}d}.
\end{equation}
An expression for $C$  as a function of our spectral parameter $s$ is needed.
In the view of the linearity of Eq. \ref{eq:RMdata}, the observed data RM can be
expressed in terms of the magnetic field $\textrm{B}$ (vector over physical
space), a response matrix $\textrm{R}$ (projecting from physical to data space)
and noise $n$ (in data space) as:  $d ={R}\,{B}+n$. From this we find
\begin{equation*}
 \begin{aligned}
C(s) &=\, \langle ({R}\,{B}+n)\, ({R}\,{B}+n)^T\rangle \,=\, {R}\, \langle
{B}\,{B}^T\rangle \, {R}^T+\langle nn^T\rangle \\
&=\,  R\, M \, R^\dagger + N = \,C_\phi(s)+ N.
 \end{aligned}
\end{equation*}
Thus to any derived Faraday rotation covariance matrix  $C_\phi= \langle \phi \, \phi^T \rangle$
an estimate for the noise covariance matrix $N= \langle nn^T\rangle$ needs to be added.

\subsection{Maximising the posterior}\label{maximazing}

The power spectrum inference within the maximum-a-posteriori approach can only be conducted numerically. Here, we maximise the logarithm of the posterior, because there are convenient analytical expressions for the gradient and
Hessian. Multiplication of the likelihood with the prior to get the posterior 
is an addition in logarithmic space, which also simplifies matters. Indices $i,j$ refer to the
degrees of freedom of the parameterisation of $s$. We find
\begin{equation}\label{eq:gradient}
\frac{\partial \ln[ P(s|d)]}{\partial s_i} =
\frac{1}{2}\mathrm{Tr}\left[ \left(dd^T-C \right)\, \left(C^{-1}\frac{\partial C}{\partial
    s_i}C^{-1}\right)\right] - \frac{1}{s_i},
\end{equation}
and if $s$ contributes only linearly to $C$:
\begin{eqnarray}\label{eq:hessian}
\frac{\partial^2 \ln[P(s|d)]}{\partial s_j \partial s_i} &=&  -\frac{1}{2}\mathrm{Tr} \left[ C^{-1} \frac{\partial C}{\partial s_j} C^{-1} \frac{\partial   C}{\partial s_i}\right] \\
&& -\mathrm{Tr}\left[\left(dd^T-C\right) \left(C^{-1}\frac{\partial C}{\partial
    s_i}C^{-1}\frac{\partial C}{\partial s_j}C^{-1}\right)\right]+\frac{\delta_{ij}}{s_i^2}.
\nonumber
\end{eqnarray}
The most probable power spectrum is then found iteratively. We use a
second-order Newton method to solve for it, with
\[s_{new}=s-\left(\frac{\partial^2 \ln[P(s|d)]}{\partial s_j \partial s_i}\right) ^{-1}\frac{\partial \ln[P(s|d)]}{\partial s_i},\]
where repeated indices are assumed to be summed over.

\subsection{The covariance matrix}\label{correlationmatrix}
An expression for the covariance matrix $C_\phi\,=\,\langle
\phi(\vec{x}_\perp) \, \phi(\vec{x}'_\perp)\rangle $
of the rotation measure map needs to be derived in the following form:
\begin{equation*}
 C_\phi(\vec{x}_\perp,\vec{x}'_\perp) = 
 a_0^2  \, \langle\,
\int_{z_1}^\infty \!\!\! dz \, n_e(\vec{x})
 B_z(\vec{x}) \int_{z_2}^\infty \!\!\! dz' \,n_e(\vec{x}')
 B_z(\vec{x}')\,\rangle .
\end{equation*}
The electron density distribution $n_e(\vec{x})$ needs to be obtained from
independent data, e.g. from X-ray observations. The magnetic field
energy is assumed to follow $n_e(\vec{x})$ according to \citet{2002A&A...387..383D}:
\begin{equation}\label{eq:magneticprofile}
{\langle B^2(\vec{x})\rangle =B^2_0\, (n_e(\vec{x})/n_0)^{2\alpha}}.
\end{equation}
The following approach is based on statistical homogeneity of the magnetic field. Therefore we
re-scale $\langle B^2(\vec{x})\rangle $ with a rescaleing function $h$,
$h(\vec{x})=\langle B^2(\vec{x})\rangle
^{0.5}/B_0=(n_e(\vec{x})/n_0)^{\alpha}$. This ensures $\langle
\widetilde{B}^2(x)\rangle =\langle B^2(\vec{x})\rangle / h^2(\vec{x})=B_0^2$ to be independent of position.
If we also assume statistical isotropy, we obtain the scaling of the
required z-component of B: $\langle B_z^2(\vec{x})\rangle
/B_{z0}^2=\langle B^2(\vec{x})\rangle /B_0^2$. Thus,
\begin{eqnarray}\label{C}
C_\phi &=& a_0^2\int_{z_1}^{\infty}\!\!\! dz\, n_e(\vec{x}_\perp,z)\, h(\vec{x}_\perp,z)\nonumber \\ 
& &
\int_{z_2}^{\infty}\!\!\! dz'\, n_e(\vec{x}_\perp',z')h(\vec{x}_\perp',z')\langle \widetilde{B}_z(\vec{x}_\perp,z)\widetilde{B}_z(\vec{x}_\perp',z')\rangle\nonumber \\
&= &  a_0^2n_0^2 
\int_{z_1}^{\infty}\!\!\! dz \, f(\vec{x}_\perp,z) \int_{z_2-z}^{\infty}\!\!\! dr_z\, f(\vec{x}_\perp',r_z+z)\, M_{zz}(\vec{r}),
\end{eqnarray}
with $\vec{r}=(\vec{x}_\perp-\vec{x}_\perp',r_z)$, where we introduced the window function $f(\vec{x})$, which is nonzero inside the Faraday screen:
\begin{equation}\label{windowfunction}
f(\vec{x})=\frac{n_e(\vec{x})}{n_0}h(\vec{x})=\left(\frac{n_e(\vec{x})}{n_0}\right)^{\alpha+1}.
\end{equation}
 $M_{zz}(\vec{r})=\langle \widetilde{B}_z(\vec{x})\widetilde{B}_z(\vec{x}+\vec{r})\rangle$ is the $zz$-part of the (rescaled) magnetic autocorrelation tensor, which we try to measure.

\subsection{Modelling the magnetic autocorrelation}\label{modelling}
$M_{zz}(\vec{r})$ can be split up into a longitudinal part $M_\mathrm{L}(r)$ and a
corresponding perpendicular part $M_\mathrm{N}(r)$, which depend only on the
absolute value of $\vec{r}$ (see  \cite{1999PhRvL..83.2957S} or \cite{2003A&A...401..835E}).
\begin{equation}\label{eq:mzz}
M_{zz}(\vec{r})=M_\mathrm{L}(r)\frac{r_z^2}{r^2}+M_\mathrm{N}(r)\frac{r_\perp^2}{r^2}
\end{equation}
The rescaled magnetic autocorrelation function\\
$\omega(r)=\langle \widetilde{B}(\vec{x})\cdot \widetilde{B}(\vec{x}+\vec{r})\rangle$ is given by
\begin{equation}\label{omega}
\omega(r) = \sum_i M_{ii}(r) = 2M_\mathrm{N}(r)+M_\mathrm{L}(r).
\end{equation}
The information of the data can be combined into a
one-dimensional correlation function and therefore does not allow us to simultaneously
determine two independent functions. Thus a relation
between $M_\mathrm{N}$ and $M_\mathrm{L}$ is necessary.
There are two suitable possibilities how to connect them.
\begin{enumerate}[a)]
\item Full isotropy assumption, which additionally assumes isotropy at
  any location in Fourier space. This yields
\begin{equation}\label{eq:mlmn}
M_\mathrm{N}(r)=M_\mathrm{L}(r).
\end{equation}
\item Considering the solenoidal character of the magnetic field by $\nabla \cdot \widetilde{B}=0$, which yields
\begin{equation}
\label{divB0}
M_\mathrm{N}(r)=\frac{1}{2r}\frac{d}{dr}(r^2 M_\mathrm{L}(r)).
\end{equation}
\end{enumerate}
The $\nabla \cdot \widetilde{B}=0$ condition leads to Eq. \ref{eq:Mij} and is more physical, but also more complicated to handle. 
Both assumptions have been implemented in \texttt{REALMAF}. 
However, the most computationally efficient way is to use the full isotropy assumption and apply the divergence relation in Fourier space, which results in a simple multiplication factor, see Section \ref{divdiscussion}.

Connecting of $M_\mathrm{L}$ and $M_\mathrm{N}$ allows us to parameterise $M_{zz}(r)$ using
\begin{equation}
M_{zz}(\vec{r})=\sum_i s_i M_{zz}^{(i)}(\vec{r}),
\end{equation}
where $M_{zz}^{(i)}(\vec{r})$ is further split into $M_\mathrm{L}^{(i)}(r)$ and $M_\mathrm{N}^{(i)}(r)$ described by Eq. \ref{eq:mzz}.
Because of the linearity of the Fourier transformation, the linear
coefficients $s_i$ in real space also appear as linear coefficients in the analogous
representation in Fourier space. 
Thus, although we are calculating the covariance matrix completely in real space, the
power spectrum can be obtained directly:
\[\omega(k)=\sum_i\, s_i\, (2M_\mathrm{N}^{(i)}(k)+M_\mathrm{L}^{(i)}(k))=\sum_i \, s_i \, w_i(k)\] .
The basis functions together with the retrieved spectral amplitudes $s_i$ can be translated into $\omega(r)$ (see Fig. \ref{spectrum} a) and the corresponding $\omega(k)$ (see Fig. \ref{spectrum} b).
The used model functions for $M_\mathrm{N}^{(i)}$ and $M_\mathrm{L}^{(i)}$ were chosen to transform into relatively well separated and well localised frequency bands $w_i(k)$ in Fourier space. 
Appendix \ref{appen1} shows the used model functions and describes how they were derived.

\subsection{The divergence relation in Fourier space}
\label{divdiscussion}
The zz-part of the magnetic autocorrelation tensor can be expressed in
Fourier space analogously to Eq. \ref{eq:mzz}
\begin{equation}\label{eq:kmzz}
M_{zz}(\vec{k})=\widehat{M}_L(k)\frac{k_z^2}{k^2}+\widehat{M}_\mathrm{N}(k)\frac{k_\perp^2}{k^2}.
\end{equation}
Applying $\nabla \cdot \widetilde{B}=0$ gives then
\begin{equation}\label{eq:kdivB0}
\widehat{M}_L(k)=0\, \Longrightarrow\, \widehat{\omega}(k)=2\widehat{M}_\mathrm{N}(k),
\end{equation}
see \cite{2003A&A...401..835E}.
The $\widehat{M}_L(k)$ from Eq. \ref{eq:kmzz} is in general not the Fourier transformation of
$M_\mathrm{L}(r)$ used in Eq. \ref{eq:mzz}. To create a connection to the real-space parameterisation $M_{zz}(\vec{r})$, we multiply $\widehat{\omega}(k)$ by 3/2, which is analogous to an addition of $\widehat{M}_L(k)$ with $\widehat{M}_L(k)=\widehat{M}_\mathrm{N}(k)$. This makes $M_{zz}(\vec{k})$ a spherically symmetric function, which forces that its Fourier back-transform $M_{zz}(\vec{r})$ must be spherically symmetric as well. Then,
\begin{equation}
M_\mathrm{L}(r)=M_\mathrm{N}(r)\, \Longleftrightarrow\, \widehat{M}_L(k)=\widehat{M}_\mathrm{N}(k),
\end{equation}
which is the full isotropy assumption defined in Eq. \ref{eq:mlmn}. Therefore, when assuming full isotropy, the divergence condition can be imposed on $\omega(k)$ to a good approximation via multiplying it with the factor 2/3. This was verified as well by numerical experiments.\footnote{Actually this relation was discovered in numerical experiments first.}
Using the full isotropy assumption reduces the processing time by up to a
factor 4 and allows a simpler spherically symmetric model for $M_{zz}$. In practice the power spectrum is multiplied with a factor
2/3 and the resulting magnetic field strength by $\sqrt{2/3}$.

Finally, the $\nabla \cdot \widetilde{B}=0$ condition, which is used by \cite{2005A&A...434...67V} as well, is not exact because of
\[\nabla \cdot \widetilde{B} = h \nabla \cdot B + B\cdot \nabla h=B\cdot \nabla h,\]
with the rescaleing function h defined in section \ref{correlationmatrix}. 
If $B$ varies on scales much smaller than the ones on which h varies, this assumption is sufficiently close to the condition $\nabla \cdot B=0$. 
Another indication that the resulting errors are negligibl small is that a totally different assumption on the statistical distribution of the magnetic fields, like full isotropy, which violates $\nabla \cdot \widetilde{B}=0$ by definition, produces only a shift by a factor of 1.5, as shown above.

\section{Application to the Hydra A data}
\label{application}

\begin{figure}[t]
  \centering
    \includegraphics[width=0.45\textwidth]{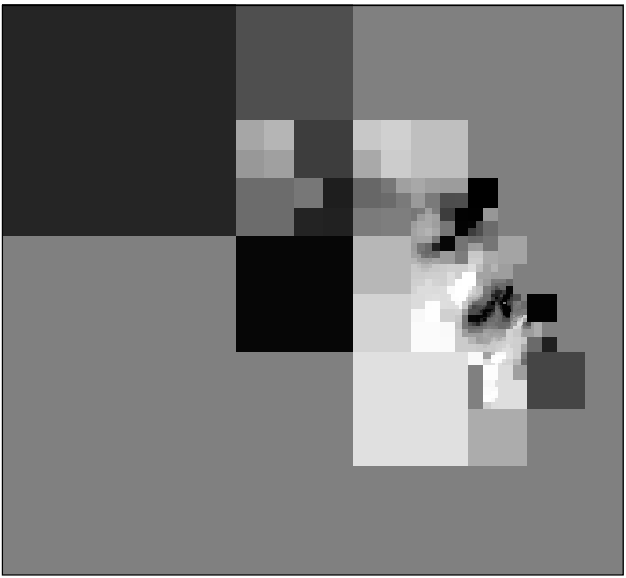}
  \caption{Merged data map of Hydra A with 1000 pixels remaining. 
The RM values range between -1700 rad/m$^2$ (dark) and 2800 rad/m$^2$ (light). 
The map has a size of about 40 kpc. 
Each square represents one value used in the analysis.}
  \label{mergeddata}
\end{figure}

\subsection{The data}\label{appldata}

\begin{figure*}[t]
 \centering
    \includegraphics[width=0.45\textwidth]{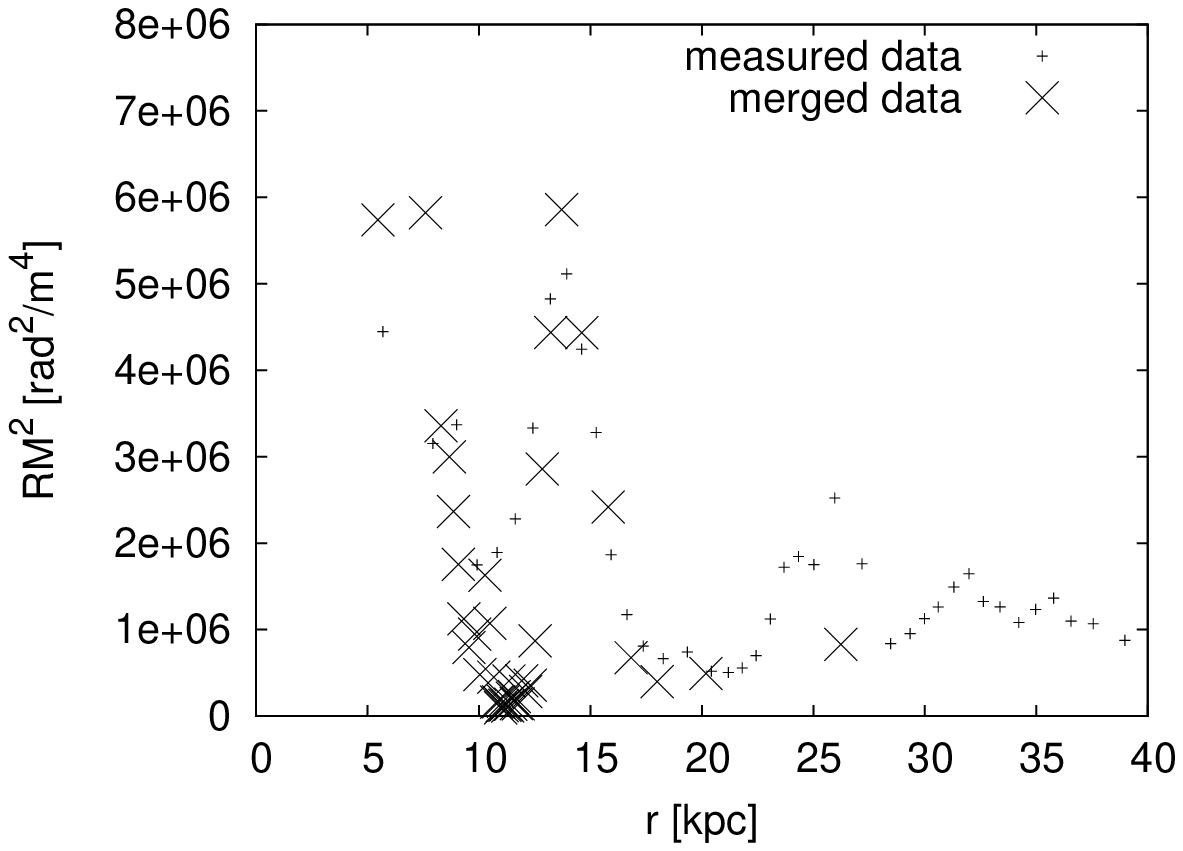}
    \includegraphics[width=0.45\textwidth]{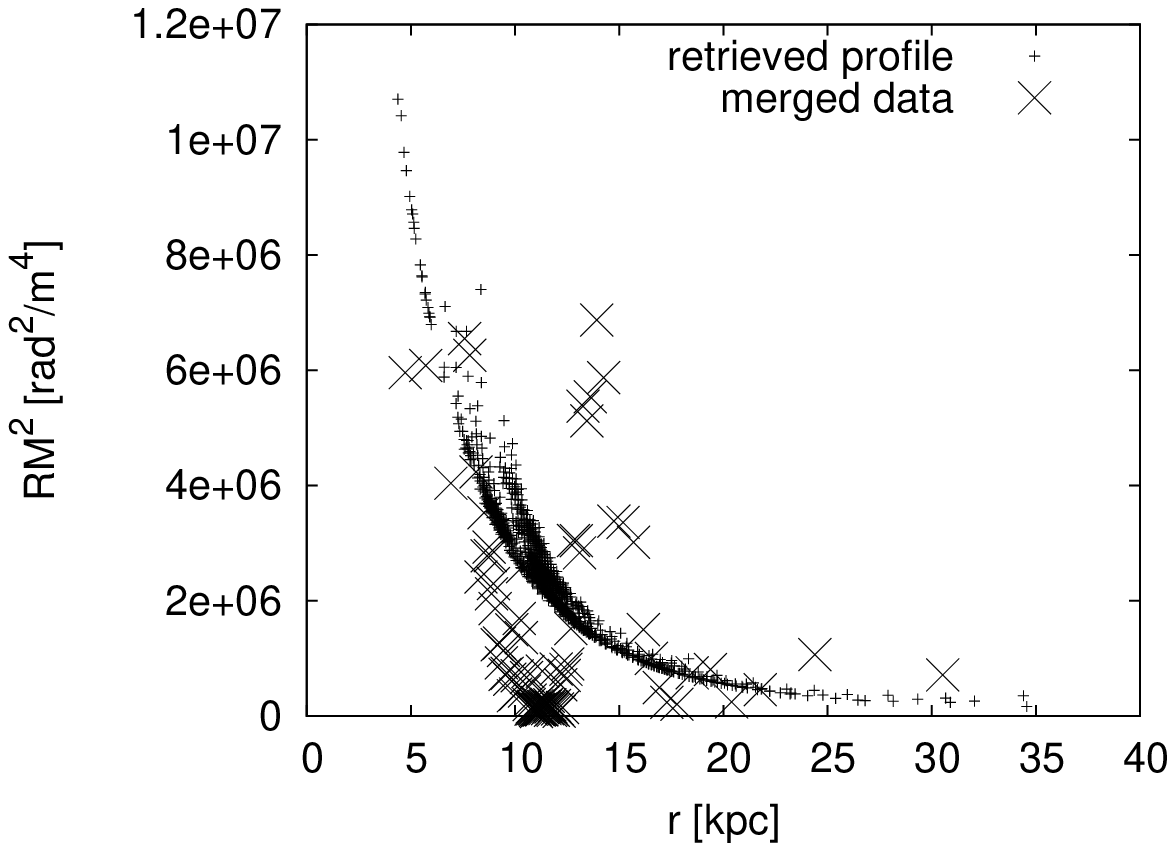}
  \caption{
Left: Radial profile of the original rotation measure data and the merged values shown in Fig. \ref{mergeddata}. 
To facilitate reading of the plot, the 1000 merged data points are combined within (variably sized) radial bins.
  \label{radialprofile_merge}
Right: Comparison of the merged profile with the theoretical radial profile of the ensemble average resulting from the analysis.
The retrieved profile was obtained from the trace of the covariance matrix described in Eq. \ref{C} and implicitly emerges from the modelled magnetic power spectrum shown in Fig. \ref{powerspectrum} and the assumed geometrical profiles of magnetic energy and electron density. 
The former determines the amplitude of the theoretical profile, whereas the latter determines the detailed radial decline. 
This theoretical curve was not obtained by a fit of the radial profile, which is shown here for illustration only. 
Instead, our inference is based on examining the cross-correlation of all pairs of points, which contain much more information compared with the autocorrelation of individual points displayed here. 
We note that the variations seen in the data with respect to the theoretical mean curve are a natural consequence of the data being only a single realisation of a random process, whereas the theoretical curve represents the ensemble average of a large number of these realisations.
}
  \label{radialprofile_compare}
\end{figure*}

High-resolution polarisation measurements of the radio lobes of Hydra A at different wavelengths, with the VLA  suitable for RM map making, were performed  by  \cite{1993ApJ...416..554T}. 
High-fidelity RM maps were constructed from these using the {\tt PACERMAN} code \citep{2005MNRAS.358..726D, 2005MNRAS.358..732V}. 
{\tt PACERMAN} uses error weighting and also non-local information to solve the $n\pi$-ambiguity and therefore produces more reliable Faraday rotation maps than algorithms that do standard fitting of Eq. \ref{eq:RMdata} pixel by pixel.
The resulting RM map was tested by \citet{2005MNRAS.358..732V} for the presence of map-making artifacts. 
These artifacts can be statistically detected using the anti-correlation of inferred RM and position angle values imprinted by misfits, as first shown by  \citet{2003ApJ...597..870E}. 
The tests show that the current RM map of the south lobe of Hydra A is prone to $n\pi$-ambiguities and other artifacts, even if generated with {\tt PACERMAN}. 
However, the north lobe, which contains lower RM values and is therefore pointing towards us, is sufficiently clean.
Therefore \citet{2005A&A...434...67V} analysed only the north lobe of Hydra A, and our application of \texttt{REALMAF} is also restricted to this area.

The resolution of the RM map is too high to take every single pixel into account in our analysis. 
A computational bottleneck is the double integral in the calculation of the model covariance matrix followed by matrix multiplications described in Section \ref{maximazing}.
To reduce the amount of required computational
resources, we merge areas in the map to bigger pixels by combining the rotation
measure values, the errors, and the pixel position in an inverse
noise-weighted fashion. 
The idea is to average the data in quadratic areas chosen so that they have roughly the same noise level.
Therefore the outer regions with lower radio surface brightness and consequently larger RM errors are more heavily averaged. 
For details on the algorithm combining the pixels, the reader is referred to  \cite{2005A&A...434...67V}.

These merged data points then form the data vector $d$ used in section \ref{basics}. Figure
\ref{mergeddata} visualises the merged map by showing the averaged values in colour within the square areas from which they were averaged. The area outside the squares does not contain any used RM values.
In Fig. \ref{radialprofile_compare} the radial profile of of RM$^2$ is shown and is compared with the ensemble averaged $\langle \mathrm{RM}^2 \rangle$ predicted by the deduced magnetic spectrum and assumed radial profile of the electron and magnetic field energy densities. 
From this figure it can be seen that the largest separation between our pixels is $r_\mathrm{max} = 33$ kpc, whereas the smallest $r_\mathrm{min} = 0.073$ kpc. 
A first guess of the optimal spectral range can be obtained based on these maximal and minimal sizes of visible structures calculated using $k_\mathrm{min} = \pi/r_\mathrm{max}$ and $k_\mathrm{max} = \pi/r_\mathrm{min}$ 
The final result in Fig. \ref{fg:powersampling} uses roughly these spectral boundaries.

Spatial averaging of RM data is not properly modelled in Eq. \ref{C}. 
In principle, it could be included into this formula and the generic inference formalism, owing to the linearity of averaging. 
However, we have to perform the double integral in Eq. \ref{C} numerically for any pair of data pixels. Adding four more integration dimensions would exceed our computational possibilities.

We accept the spectral error resulting from the data averaging operation because pixels that resulted from averaging over large regions are automatically only contributing information about scales above their size. 
Any baseline formed by them with other pixels will be of the order of the pixel size or larger.
The product of the RM values of a pair of pixels is compared with the expected data correlation over this baseline distance. 
Thus these pixels determine the inferred spectrum mostly on scales on which their internal RM substructure, which was averaged out, does not matter. 

The silhouette of the radio lobe used to probe the Faraday screen breaks the translational invariance of our problem, in addition to the effect of the electron density profile. 
However, this does not impact directly our data covariance matrix $C$, because this only contains the expected (averaged over an ensemble of data realisations) correlation of individual pairs of positions in the RM map. 
The silhouette restricts the available positions and therefore the number of baselines, which means that on larger scales, we have less information. 
However, the Bayesian formalism is informed about this and just increases the uncertainty margins for poorly constrained data modes. 

We also require to specify the noise covariance matrix. 
Assuming the noise at different pixels to be uncorrelated,
the noise matrix is diagonal. The diagonal elements can
be approximated by the square of the RM-deviation of each pixel given by the error map produced by \texttt{PACERMAN}. 

The used rotation measure map has a mean value of about
$600 \,{\textrm{rad}}/{m^2}$. The galactic rotation measure at the
position of Hydra A is about 1 percent of this and therefore
negligible, see \cite{2004mim..proc...13J}. Therefore we assume that the mean value is
caused by cosmic variance which is in turn processing a too small sample of the
cluster. This mean value is correctly interpreted by our
method in that it adds power to the small $k$ part of the magnetic power spectrum,
compared with the case in which we subtract the mean value from the data $d$.

\subsection{Window function}\label{applwindow}
\begin{figure}[t]
  \centering
    \includegraphics[width=0.45\textwidth]{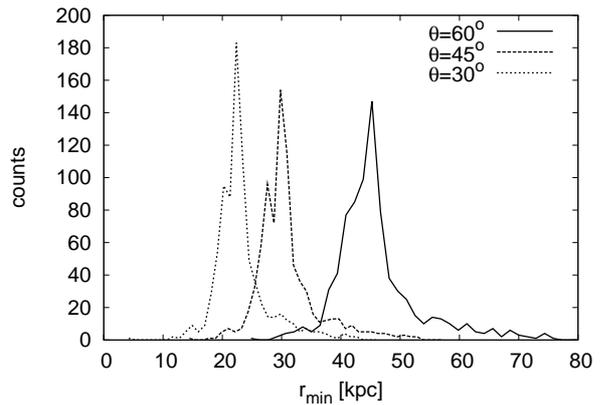}
  \caption{Histogram of the minimal distance $R_{min}$ from the cluster
    centre for 800 data points depending on the assumed
    projection angle $\theta$. Magnetic fields at $r < R_{min}$ are
    invisible to the method.}
\label{histogram}
\end{figure}

The window function $f$ defined in Eq. \ref{windowfunction} requires the electron density distribution $n_e(x)$. Hydra A is a cooling-core cluster. Therefore we assume a spherically symmetric double beta profile defined as
\begin{equation}\label{eq:windowfunction}
{n_e(x)= \left[ n_{e1}^2 \left(1+ \left( \frac{r}{r_{c1}} \right)^2 \right)^{-3\beta} + n_{e2}^2 \left( 1+\left( \frac{r}{r_{c2}}\right)^2 \right)^{-3\beta} \right]^{1/2}},
\end{equation}
with $n_{e1}=0.056\,\textrm{cm}^{-3}$, $n_{e2}=0.0063\,\textrm{cm}^{-3}$,
$r_{c1}=33.3\,\textrm{kpc}$, $r_{c2}=169\,\textrm{kpc}$ and
$\beta=0.766$, as fitted by \cite{2005A&A...434...67V} using data from \cite{1999ApJ...517..627M}. 
Various works propose that the observed radio lobes of active
galactic nuclei, like Hydra A, form cavities in the cluster
atmosphere, see e.g. \cite{1993MNRAS.264L..25B}, \cite{1999dtrp.conf..275E}, \cite{2000ApJ...534L.135M}, \cite{2000MNRAS.318L..65F}, and  \cite{2001ApJ...557..546D}. In these cavities the polarised radio light is produced, but $n_e(x)$ is presumably so low that no significant Faraday rotation is expected to occur there. 
The fact $n_e(x)=0$ for
any $x \in \,\textrm{cavity}$ can be taken into account by adjusting
the integration boundaries in Eq. \ref{C}. In order to calculate the
important lower integration boundary, where the Faraday active area
starts, we need to know the geometry of the cavity. The cavity
in our used map-section can be approximated by an ellipse with the
projected distance 24.9 kpc from cluster centre, projected semimajor
axis 20.5 kpc, and semiminor axis with 12.4 kpc, as
\cite{2007ApJ...659.1153W} propose.
Additionally the Hydra A north lobe is tilted by an angle $\theta$
towards the observer. Depending on $\theta$ the innermost radius of the probed Faraday
active area shifts, as can be seen in Fig. \ref{histogram}. There it becomes apparent
that the sensitivity of our analysis starts at radii of about 20 to 30 kpc from the cluster centre depending on $\theta$.
$\theta$ is not well determined and its estimates range from about $30^o$ (\cite{1993ApJ...416..554T}, if the cavity effect is added) to more than $45^o$ \citep{2008MNRAS.391..521L} to $60^o$ \citep{2007ApJ...659.1153W}.

\subsection{Measuring the magnetic profile}\label{alphameasuring}

\begin{figure}[t]
  \centering
    \includegraphics[width=0.45\textwidth]{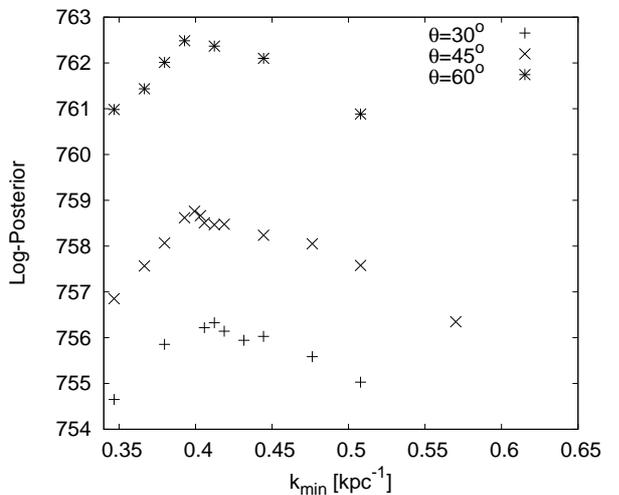}
  \caption{Sampling of $k_{min}$ for given $\alpha=1$}
\label{fg:kminsampling}
\end{figure}

The index $\alpha$ of the magnetic field can be retrieved from the
data. In contrast to the power spectrum\footnote{and the directly obtained central magnetic field strength $B_0$}, which is built dynamically by our algorithm, $\alpha$ has to be set up for every processing
run. Maximising the posterior with respect to $\alpha$ provides indications on most probable
configurations. One drawback is that $\alpha$ also depends on the
low-frequency cut-off $k_{min}$ (position of the top of the first
spectral bin) chosen; meaning a two-dimensional problem needs to be optimised. The result we find is $\alpha=1$ with $k_{min}=0.42\, \textrm{kpc}^{-1}$. 
Because this result only depends on the large scales, a rather low-resolution study with 300 data points was sufficient and permitted a fast investigation of the 2-d parameter space. Sampling of $k_{min}$ for three different projection angles $\theta$ can be found in Fig. \ref{fg:kminsampling}.
\begin{figure}[t]
  \centering
    \includegraphics[width=0.45\textwidth]{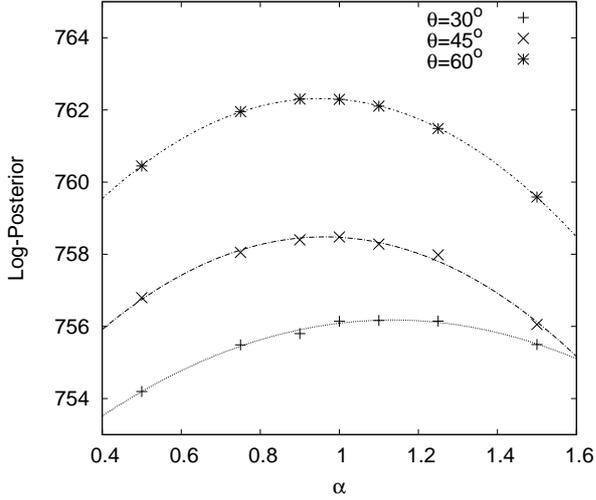}
  \caption{Sampling of $\alpha$ for given $k_{min}=0.42\, \textrm{kpc}^{-1}$}
\label{fg:alphasampling}
\end{figure}
Cut-offs with $k<0.42\, \textrm{kpc}^{-1}$ correspond to $\alpha>1$
and vice versa. Smaller $\theta$ implies a smaller physical distance
between the data points, which results in a small shift of $k_{min}$. However, it is negligible and inside the statistical variance of about $0.05\, \textrm{kpc}^{-1}$.
Fig. \ref{fg:alphasampling} shows a consistency check by sampling $\alpha$ for the given $k_{min}=0.42\, \textrm{kpc}^{-1}$. The curvature in Fig. \ref{fg:alphasampling} gives roughly a 1-sigma variance for $\alpha$ of about $\Delta \alpha = \pm 0.3$, while assuming Gaussianity around the peak. 
When we also consider the impact of the variance of $k_{min}$, we find $\alpha=1 \pm 0.5$.

Surprisingly the posterior maximum is rising for rising $\theta$ with no apparent
limit. 
We suppose that this has nothing to do with the probability of the
physical model and therefore $\theta$ cannot be restricted using the data of the Hydra A north lobe only. 
Indeed, $\theta$ is a parameter, that only affects the
integration bounds of Eq. \ref{C}, whereas the power spectrum, the magnetic scaling $\alpha$, and the cut-offs affect the integrand.

\subsection{Power spectrum and magnetic characteristics}

\begin{figure}[t]
  \centering
    \includegraphics[width=0.45\textwidth]{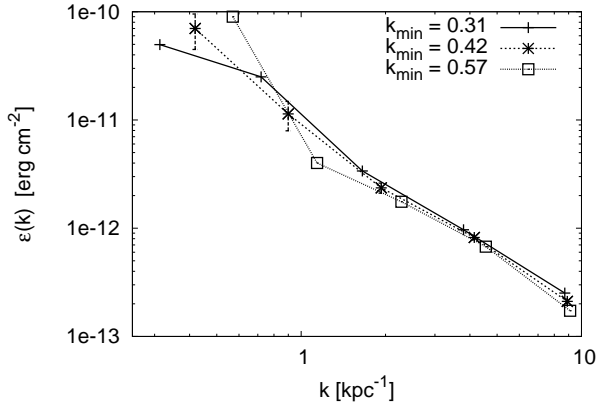}
  \caption{Power spectrum with an optimal, a too small and a too large cut-off $k_{min}$}
\label{fg:powersampling}
\end{figure}
\begin{figure}[t]
  \centering
    \includegraphics[width=0.45\textwidth]{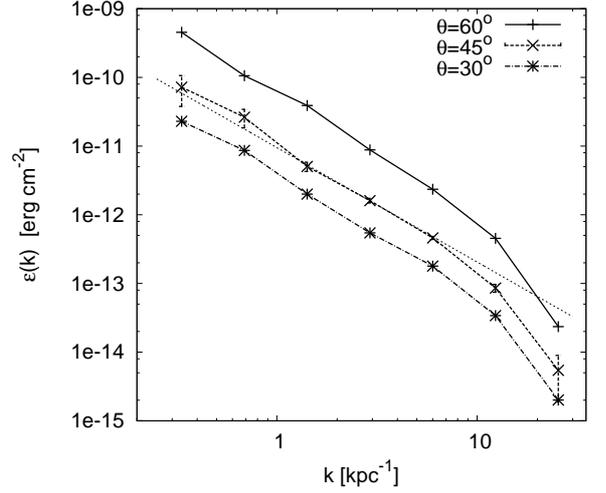}
  \caption{Inferred power spectrum for three assumed projection angles and comparison with a Kolmogorov slope (dotted). Note that for $k>10\,\mathrm{kpc}^{-1}$ beam effects reduce the amount of power and are probably responsible for the steepening.}
\label{powerspectrum}
\end{figure}

The shape of the power spectrum depends on the spectral cut-offs
$k_{min}$ and $k_{max}$. We retrieved the most probable $k_{min}$ to
be 0.42 kpc$^{-1}$ in Section \ref{alphameasuring}. Obviously, the most
probable configuration is a power law with no large-scale turnover visible within 
the probed spatial scales, as can be seen in Fig. \ref{fg:powersampling}. 
The decline for smaller $k_{min}$
may be caused by the limited rotation measure map dimensions and the down-pull of thereby unconstrained amplitudes by Jeffreys prior. A larger
$k_{min}$ will remove necessary degrees of freedom of the model at
the large scales, which forces the algorithm to deposit additional power at the
neighbouring scale.
The high-frequency cut-off is limited by the resolution of the map and instrumental beam. A beam with spatial scales of $0.32\, \textrm{kpc}$ (FWHM) gives $k_{max}=10\, \textrm{kpc}^{-1}$. 
Below, we increase the resolution by increasing the number of data points to 1000 and extend the spectrum to a maximal possible range. 
We made three processing runs with $\alpha=1$ and three possible projection angles, see Fig. \ref{powerspectrum} and Table \ref{characteristics}. To investigate the dependence of the results on $\alpha$ we made two more runs with $\alpha=0.5$ and $\alpha=1.5$, see Table \ref{characteristics2}. Formulas for the calculation of the magnetic characteristics are described in Appendix \ref{charhowto}. The given statistical uncertainties refer to static $\alpha$ and $\theta$. These errors are propagated from the error bars of the power spectrum.

The power law ranges from scales of 0.3 kpc (beam FWHM) to about 8 kpc. The fall-off in power at small scales is caused by instrumental beam effects. 
However, the larger scales are not greatly affected by the
beam. To measure the spectral index we ignore the first spectral bin, which is 
possibly affected by map boundary limits as well as the last two beam
affected bins. 
The dotted line in Fig. \ref{powerspectrum} represents
a Kolmogorov slope of $\epsilon(k)\propto k^{-1.67}$ compared with the plot using $\theta=45^o$. 
We achieve a
perfect fit for the central part of the retrieved power spectrum,
which is also inside the larger statistical variance of the parts at
smaller k. There is a noticeable dependence of the spectral index on
$\alpha$ with minimal and maximal slopes of 1.5 and 2.1, which should
be regarded as improbable worst cases. $\theta$ has a big impact on
the resulting magnetic field strengths. 
Assuming a larger $\theta$
implies a shorter interaction path with the magnetic field and
therefore greater necessary magnetic field strengths. The derived $B_0$
in the cluster centre assumes that the magnetic profile given by
Eq. \ref{eq:magneticprofile} can be extrapolated with a constant
$\alpha$ into the cluster centre. However, the sensitivity starts at
regions from about 30 kpc from the cluster centre, see Fig. \ref{histogram}, and
a flattening of the magnetic profile at smaller radii is well possible. $B_{50}$
gives the directly probed magnetic field strengths at 50 kpc
distance.

The impact of the instrumental beam on the calculated magnetic field
strength is low. A continuation of the power law to the small scales would
give an additional $0.2\,\mu$G at most.

\begin{table}[t]
\centering
\begin{tabular}{c  c c c c c}
\hline\hline
 $\theta$ & $B_0$ [$\mu$G] & $B_{50}$ [$\mu$G] & $\lambda_B$ [kpc]& spectral index \\
\hline
$30^o$ & 21 $\pm$ 1 & 9.3 & 5.0 $\pm$ 1.0 & 1.70 $\pm$ 0.14\\
$45^o$ & 36 $\pm$ 2 & 16 & 5.2 $\pm$ 1.1 & 1.73 $\pm$ 0.13\\
$60^o$ & 85 $\pm$ 5 & 37 & 5.3 $\pm$ 1.2 & 1.85 $\pm$ 0.14\\
\hline
\end{tabular}
\caption{Magnetic field characteristics for most probable $\alpha=1$}
\label{characteristics}
\end{table}

\begin{table}[t]
\centering
\begin{tabular}{c c c c c c}
\hline\hline
 $\alpha$ & $B_0$ [$\mu$G] & $B_{50}$ [$\mu$G] & $\lambda_B$ [kpc]& spectral index \\
\hline
$0.5$ & 18 $\pm$ 1 & 12 & 4.8 $\pm$ 1.1 & 1.56 $\pm$ 0.14\\
$1.5$ & 70 $\pm$ 4 & 21 & 5.5 $\pm$ 1.2 & 2.06 $\pm$ 0.14\\
\hline
\end{tabular}
\caption{Magnetic field characteristics for $\theta=45^o$ and maximal and minimal possible $\alpha$}
\label{characteristics2}
\end{table}

\section{Comparison with other methods}\label{comparison}

\begin{table}[t]
\centering
\begin{tabular}{c c c c c}
\hline\hline
author & $\alpha$ & $B_0$ in $\mu$G & spectral index & large-scale turnover \\
\hline
Kuchar & 1.0 & 36 & 1.73 & no \\
Laing & 0.25 & 19 & 0.8 & no \\
Vogt & 0.5 & 7.3 & 1.67 & yes\\
\hline
\end{tabular}
\caption{Recent results of Hydra A for $\theta=45^o$}
\label{other results}
\end{table}

The magnetic power spectrum of Hydra A was previously measured by
\cite{2005A&A...434...67V} and by \cite{2008MNRAS.391..521L}. Both
find significantly lower central magnetic field strengths than we
do. This is because of a much flatter magnetic field profile inferred characterised by a smaller index
$\alpha$. Moreover \cite{2005A&A...434...67V} found a bump, a cut-off at spatial scales of
about 2 kpc and more, but the spectral index for small scales is still
similar to our current results. However, Laing finds a much flatter spectral index of
about 0.8. Table \ref{other results} compares the results for a
projection angle of $\theta=45^o$.

\begin{figure}[t]
  \centering
    \includegraphics[width=0.45\textwidth]{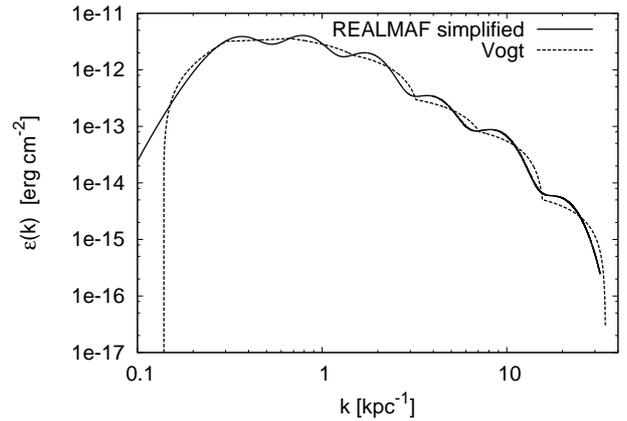}
  \caption{Reconstruction of the results of the Fourier space method
    by \cite{2005A&A...434...67V} by adding the removed
    simplifications for $\alpha=0.5$ and $\theta=45^o$. This plot
    illustrates the continuous result taking the model for the power
    spectrum into account.}
\label{f-consistence}
\end{figure}

\begin{figure}[t]
  \centering
    \includegraphics[width=0.45\textwidth]{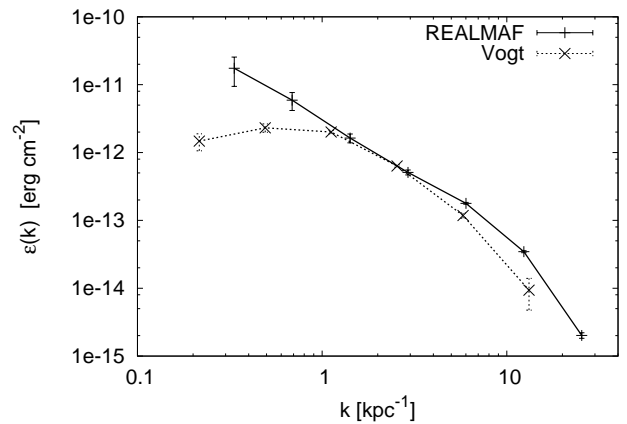}
  \caption{Power spectrum using the \texttt{REALMAF} and the method by \cite{2005A&A...434...67V} for $\alpha=0.5$ and $\theta=45^o$. According to Fig. \ref{f-consistence} the Fourier method is equivalent to explicitly adding simplifications to the real space method.}
\label{f-difference}
\end{figure}

To partly reconstruct the results of \cite{2005A&A...434...67V}, we applied the two
simplifications they have used in the calculation of the covariance matrix in
Eq. \ref{C},
\begin{enumerate}
\item extending the second lower integration bound to
-$\infty$
\item merging the two window functions $f$ assuming the magnetic
autocorrelation changes on much smaller scales than the window
function.
\end{enumerate}
Then we compared the results with the Fourier method of Vogt,
but adding noise effects, the cavity structure, and a more advanced triangular
parameterisation of the power spectrum, see Fig. \ref{f-consistence}. 
We reconstructed the spectral bump found by \cite{2005A&A...434...67V} and found a perfect match between both methods. Figure \ref{f-difference} shows the difference to our method
without these simplifications.

\begin{figure}[t]
  \centering
    \includegraphics[width=0.45\textwidth]{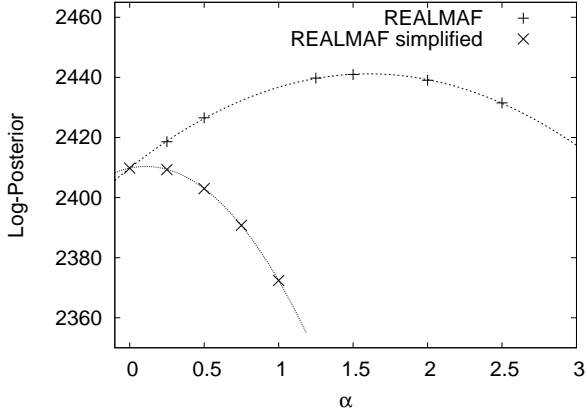}
  \caption{Sampling of $\alpha$ with and without simplifications used
    in \cite{2005A&A...434...67V} for $\theta=45^o$ and $k_{min}=0.3\,
  \textrm{kpc}^{-1}$}
\label{f-alphasimple}
\end{figure}

Consistent with \cite{2005A&A...434...67V}, the most likely $\alpha$ is
shifted to much lower values if one adopts these
simplifications, as can be seen in Fig. \ref{f-alphasimple}. 
However, instead of a falling plateau of the posterior between
$\alpha=0.1$ and $\alpha=0.8$ and a cut-off for the unphysical range $\alpha<0.1$ we
again find a Gaussian distribution, but with a maximum at
0.1. This very low $\alpha$ value indicates that the bump found by
\cite{2005A&A...434...67V} may be explained by a power leak on large scales produced by an implicit increase of the large-scale
contribution of the window function.

One of the major simplifications used by \cite{2008MNRAS.391..521L} is a linear and position-independent relation
between the RM power spectrum and the magnetic
autocorrelation\footnote{applied by \cite{2003A&A...412..373V} as well}, which
is only valid for a deep and statistically homogeneous Faraday screen. Both assumptions are not
very well met for the Hydra A cluster. 
If we simulate a constant window function by setting $\alpha=-1$ (for which the effective window function, as given by Eq. \ref{windowfunction}, becomes flat) and use symmetric integration bounds in
Eq. \ref{C} we get a spectral index of about
1.1. For $\alpha=0.25$ we get a central magnetic field strength of 12.9
$\mu$G, which is much closer to their results.

The remaining difference to the spectral index of 0.8 and field strength of 19
$\mu$G may be partly caused by the beam correction applied by \cite{2008MNRAS.391..521L}, which
lifts the high-frequency part of the spectrum, which is possibly contaminated with noise residuals from the RM-map-making algorithms. An implementation of a beam correction as a convolution in
real space is not feasible in \texttt{REALMAF}, because it would be computationally too
expensive and an implementation in Fourier space is in general only
possible with simplifying assumptions. 
However, the exact frequency-dependent instrumental beam is unknown anyway, and we believe its effect can be neglected at scales larger than the beam size. 

\section{Conclusions}
\subsection{Summary}
We developed a new method to retrieve magnetic power spectra from
Faraday rotation maps implemented in the \texttt{REALMAF} (REAL
MAgnetic Fields) code, where we modelled the magnetic autocorrelation
function in real space.

Furthermore we introduced and verified a way
to take the solenoidal character of the magnetic fields into account
by using a simple full spherical model for the magnetic
autocorrelation and multiplying with a proportionality factor. In
contrast to the method by \cite{2005A&A...434...67V} with a model in
Fourier space, this permits us to alleviate some previously required simplifications in
the processing. However, we are able to reconstruct their results by
synthetically adding their simplifications to our method.

When we apply {\tt REALMAF} to the north lobe of the Hydra A galaxy cluster, we find a
power law power spectrum in the spatial range of 0.3 kpc to 8 kpc, with no visible large-scale turnover within this range, as reported by \cite{2005A&A...434...67V}. 
The magnetic density profile, which we modelled with an index $\alpha$ applied to the electron density profile, seems to follow the electron density profile with $\alpha=1\pm 0.5$. 
The somewhat large uncertainty on  $\alpha$ also affects the spectral index of the power law, which is then in the range of 1.5 to 2.1. 
The exact projection angle of the used radio lobe Hydra A north is not well restricted, but it only seriously affects the absolute value of the magnetic energy density in the cluster and not the spectral slope. 
Depending on the angle we obtain field strengths ranging from 20 $\mu$G up to 80 $\mu$ G in the cluster centre. 
However, these are extrapolations, because the measurement is only sensitive at a distance of about 50 kpc from cluster centre, where we find fields between 10 $\mu$G and 30 $\mu$G. 
The extrapolation of the model into the centre may be questionable.

Theories describing the generation of magnetic fields by turbulence like \cite{2006MNRAS.366.1437S},
\cite{2006A&A...453..447E} and \cite{2006PhPl...13e6501S} forecast a
large-scale turnover, which might be visible inside our spectral
range. Compared with \cite{2005A&A...434...67V} we found no large-scale turnover within the spectral range up to 8 kpc. 
A turnover at larger scales is, however, not excluded by our analysis.

\subsection{Outlook}
The \texttt{REALMAF} code will permit us to analyse the Faraday rotation data of other galaxy clusters. This will hopefully allow us to study magnetic field spectra over a large range of spatial scales and in a variety of cluster environments. A usable map of Hydra A south would further increase the probed range at large scales and a map with a higher resolution would increase the range at small scales. We published the \texttt{REALMAF} code for general scientific usage under an open source license\footnote{{\tt http://sourceforge.net/projects/realmaf/}}. 

Upcoming telescopes such as EVLA, LOFAR, LWA, ASKAP, and SKA will certainly
provide the necessary high-fidelity Faraday rotation data of many more
galaxy clusters for further studies with \texttt{REALMAF}. This may help to figure out the origin of cluster magnetic fields and the role they play in the metabolism of the intra-cluster medium.

\begin{acknowledgements}
Special thanks to Corina Vogt, who made her code used in
\cite{2005A&A...434...67V} available for insight, 
thanks to Daria Guidetti and Robert Laing for discussions and to an anonymous referee for constructive comments, and thanks to Henrik Junklewitz and Niels Oppermann for comments on the manuscript.
This research was performed in the framework of the DFG Forschergruppe 1254 ''Magnetisation of Interstellar and Intergalactic Media: The Prospects of Low-Frequency Radio Observations''.
\end{acknowledgements}

\appendix
\section{Model functions}
\label{appen1}
\begin{figure}[t]
  \centering
    \includegraphics[width=0.45\textwidth]{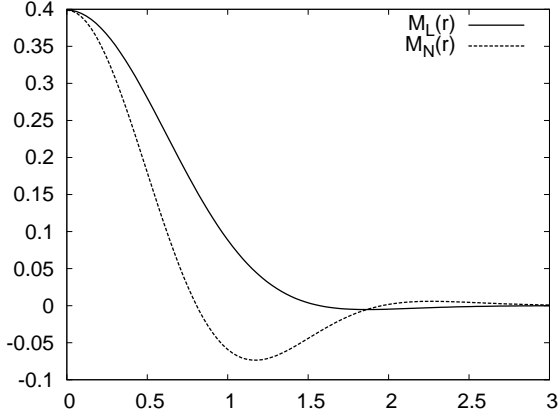}
  \caption{model functions $M_\mathrm{N}^{(i)}(r)$ and $M_\mathrm{L}^{(i)}(r)$ with $a_i=\frac{3}{4}$ and $b_i=1$}
  \label{real}
\end{figure}

\begin{figure}[t]
  \centering
    \includegraphics[width=0.45\textwidth]{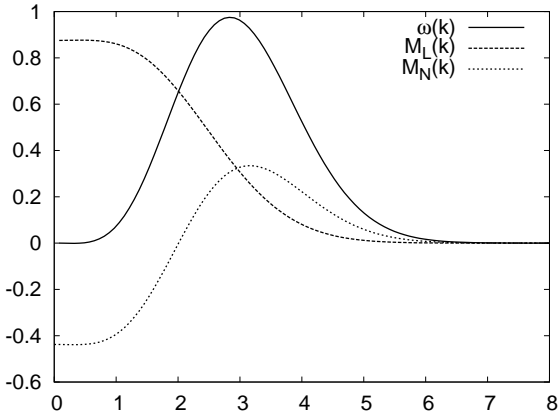}
  \caption{Model functions $M_\mathrm{N}^{(i)}(k)$, $M_\mathrm{L}^{(i)}(k)$ and $\omega^{(i)}(k)$ in Fourier space with $a_i=\frac{3}{4}$ and $b_i=1$}
  \label{fourier}
\end{figure}

Here we describe the construction of our spectral model functions.
We start with an ansatz for $M_\mathrm{L}(r)$ for the case $\nabla \cdot
\widetilde{B}=0$ and use a Gaussian multiplied with a cosine. The corresponding
$M_\mathrm{N}(r)$ is obtained by Equation \ref{divB0}. Thus we adopt
\begin{equation}\label{M(r)}
   \begin{aligned}
M_\mathrm{L}^{(i)}(r) =&
\frac{1}{\sqrt{2\pi}}b_i^3\exp[-\frac{1}{2}\frac{r^2}{a_i^2}]\cos[b_ir]\,,\,\textrm{and}\\
M_\mathrm{N}^{(i)}(r) =& \frac{b_i^3}{2a_i^2
  \sqrt{2\pi}}\exp[{-\frac{r^2}{2a_i^2}}]((2a_i^2-r^2)\cos{[b_i
  r]}\\
+& a_i^2 b_i r \sin{[b_i r]}).
\end{aligned}
\end{equation}
These functions are Fourier transformed analytically using
\[M(k)=4\pi\int_0^\infty r^2 M(r)\frac{\sin{[k r]}}{k r}dr,\]
see \cite{2003A&A...401..835E}. We
bind the free parameter $a_i$ as
\[a_i=\frac{3}{4}b_i,\]
 because in this case $\omega^{(i)}(k)=2M_\mathrm{N}^{(i)}(k)+M_\mathrm{L}^{(i)}(k)$ is fully positive and concentrated around a
maximum at $k=3\, b_i$. This allows us to put the power spectrum together by
relatively well separated and located frequency bands,
as Fig. \ref{fourier} illustrates. Fig. \ref{real} shows the model
functions in position space. Especially $M_\mathrm{N}(r)$ has a strong negative part, because
any closed field line has to come somewhere back through the plane
perpendicular to it.

If we assume full isotropy, $M_{zz}(r)$ becomes spherically symmetric. In this case a suitable model function can be obtained by Eq. \ref{M(r)}:
\begin{equation}
M_{zz}^{(i)}(r)=\frac{1}{3}\omega^{(i)}(r)=\frac{1}{3}(M_\mathrm{L}^{(i)}(r)+2M_\mathrm{N}^{(i)}(r)).
\end{equation}
This implicitly assumes $M_\mathrm{N}(r)=M_\mathrm{L}(r)$. Thus we get
\begin{enumerate}[\bfseries a)]
\item \textbf{using $\nabla \cdot \widetilde{B}=0$}

in position space:
\begin{equation}
   \begin{aligned}
M_\mathrm{L}^{(i)}(r)&=\frac{b_i^3}{\sqrt{2\pi}}\exp[-\frac{8}{9} b_i^2 r^2] \cos[b_i r]\\
M_\mathrm{N}^{(i)}(r)&=
\frac{b_i^3}{18\sqrt{2\pi}}\exp[-\frac{8}{9}b_i^2r^2](2(9-8b_i^2r^2)\cos[b_ir]\\
-& 9b_ir\sin[b_ir]),
\end{aligned}
\end{equation}
in Fourier space:

\begin{equation}
   \begin{aligned}
M_\mathrm{L}^{(i)}(k)=&\frac{27}{64}\pi\left(\frac{k-b_i}{k}\exp[-\frac{9(b_i-k)^2}{32b_i^2}]\right.\\
+&\left.\frac{k+b_i}{k}\exp[-\frac{9(b_i+k)^2}{32b_i^2}]\right)\\
M_\mathrm{N}^{(i)}(k)=&\frac{27}{128}\pi\left((\frac{9(b_i-k)^2}{16b_i^2}-1)\exp[-\frac{9(b_i-k)^2}{32b^2}]\right.\\
&+\left. (\frac{9(b_i+k)^2}{16b_i^2}-1)\exp[-\frac{9(b_i+k)^2}{32b_i^2}]\right),
\end{aligned}
\end{equation}

and
\item \textbf{using full isotropy}

in position space:

\begin{equation}
   \begin{aligned}
M_{zz}^{(i)}(r)=&\frac{b_i^3}{27\sqrt{2\pi}}\exp[-\frac{8}{9}b_i^2r^2]\\
& \left((27-16b_i^2r^2)\cos[b_ir]-9b_ir\sin[b_ir]\right),
   \end{aligned}
\end{equation}

in Fourier space:

\begin{equation}
   \begin{aligned}
M_{zz}^{(i)}(k)=&\frac{27}{1024b_i^2k}\pi\exp[-\frac{9(b_i+k)^2}{32b_i^2}]\\
&\left(16b_i^3+9b_i^2k+18b_ik^2+9k^2+9k^3\right.\\
&+ \left.\exp[\frac{9k}{8b_i}](-16b_i^3+9b_i^2k-18b_ik^2+9k^3)\right).
\end{aligned}
\end{equation}

\end{enumerate}

\section{Calculation of magnetic the field characteristics}\label{charhowto}

Here we provide the used formulae for the magnetic field characteristics.
\begin{enumerate}[\bfseries a)]

\item \textbf{Magnetic field strength}

The magnetic energy density can be calculated by integrating over the power spectrum or analogously taking the autocorrelation function in real space at the origin $\omega(0)$. The magnetic field strength in the centre of the cluster, where $f=1$ and therefore $\widetilde{B}=B$, is then
\begin{equation}
B  =  \sqrt{\omega(0)}.
\end{equation}

\item \textbf{Magnetic autocorrelation length}
\begin{equation}
\lambda_{B}=\frac{\int_{-\infty}^\infty dr \omega(r)}{\omega(0)}.
\end{equation}

\item \textbf{1-dimensional power spectrum}

Transformation from the three-dimensional power spectrum $\omega (k)$ to
the one-dimensional $\epsilon (k)$ is done using \citep{2003A&A...401..835E}
\begin{equation}\label{eq:dimensions}
\epsilon(k)=\frac{k^2\omega(k)}{2(2\pi)^3}.
\end{equation}

\item \textbf{Spectral index using a Bayesian approach}

The spectral index can be calculated from the slope of the power spectrum in logarithmic scale. The fitting formula is:
\[\epsilon(k)=\epsilon_0(\frac{k}{k_0})^{-r}=q k^{-r}\]
The variance from the above power law is then defined as:
\[\delta s=s_i-q k_i^{-r}\]
The allowed statistical deviations of s are imprinted in a covariance
matrix $D_{ij}=\langle \delta s_i\, \delta s_j\rangle$, which is automatically obtained from the inverse of the Hessian matrix
calculated with the maximum a posteriori approach as described in
Section \ref{maximazing}. The probabilities of the spectral amplitudes
s of the power spectrum can be approximated as a Gaussian, where
we can neglect the normalisation factor, because it is independent of
q and r:
\[P(\delta s|q,r)\propto e^{-\frac{1}{2}\delta s^T D^{-1}
  \delta s}\].
Only a part of the power spectrum is used to fit the power law. $D_{ij}$ is then
the projection (cut) of the full inverse Hessian matrix
$\widetilde{D}_{ij}$ obtained from Eq. \ref{eq:hessian}
\[D_{ij}=<\delta s_i \delta s_j> = \psi^T\widetilde{D}_{ij}\psi,\]
with the projection operator $\psi=\{0,...,1,...1,...,0\}$ selecting the used spectral bins for
the fit. 
Using the Bayesian theorem from Eq. \ref{eq:bayesiantheorem}:
\[P(q,r|\delta s)\propto P(\delta s|q,r),\]
where we already applied a uniform prior distribution $P(r,
q)=\textrm{const}$. $P(q,r|\delta s)$ has to be maximised to achieve the desired slope r within a maximum-a-posteriori approach. To numerically find the optimum, the gradients and the 2x2 Hessian matrix are necessary:
\begin{eqnarray*}
 \frac{\partial \ln P}{\partial r} &=&  -q\ln[k_i]k_i^{-r}D_{ij}^{-1}(s_j-qk_j^{-r})\\
\frac{\partial \ln P}{\partial q} &=&
k_i^{-r}H_{ij}^{-1}(s_j-qk_j^{-r})\\
 \frac{\partial^2 \ln P}{\partial q^2} &=&  -k_i^{-r}D_{ij}^{-1}k_j^{-r}\\
 \frac{\partial^2 \ln P}{\partial r^2} &=&
 q(q\ln[k_i]k_i^{-r}D_{ij}^{-1}\ln[k_j]k_j^{-r}\\
&&-(\ln[k_i])^2k_i^{-r}D_{ij}^{-1}(s_j-qk_j^{-q}))\\
\frac{\partial^2 \ln P}{\partial q \partial r} &=& 2q\ln[k_i]k_i^{-r}D_{ij}^{-1}k_j^{-r}-\ln[k_i]k_i^{-r}D_{ij}^{-1}s_j .
\end{eqnarray*}
An expression for the second unknown $q$ can be found analytically. If the initial value of $r$ is given, the corresponding initial value $q$ can be calculated easily. This is necessary because the first guess must be close to the maximum to attain convergence.
\begin{eqnarray*}
q=\frac{k_i^{-r}D_{ij}^{-1}s_j}{k_i^{-r}D_{ij}^{-1}k_j^{-r}}
\end{eqnarray*}

\end{enumerate}

\bibliographystyle{aa}
%\bibliography{powerhydra}

\end{document}